\documentclass[11pt]{article}
\usepackage[pdftex]{graphicx}
\usepackage{hyperref}
\usepackage{float} % to place the table, image on the exact place where required, with comment \begin{figure}[H] or \begin{table}[H]
\usepackage{multirow}% to adjust the number of column and row
\hypersetup{
	colorlinks   = true, %Colours links instead of ugly boxes
	urlcolor     = blue, %Colour for external hyperlinks
	linkcolor    = blue, %Colour of internal links
	citecolor   = red %Colour of citations
}

\usepackage{cite}
\usepackage{multirow}
\usepackage{amsmath,amsfonts,amssymb,amsthm,nccmath,latexsym,mathtools}
\usepackage{xcolor}
\usepackage{amsmath}

\usepackage{tikz,lipsum,lmodern} %these are for boxes
\usepackage[]{tcolorbox}
\definecolor{light-gray}{gray}{0.95}

\usepackage{dsfont}
\usepackage{cancel} %to strike out obliquely
\usepackage{soul}
\usepackage{ amssymb }
\usepackage{footnote}

\def\no{\nonumber}

\def\d{\delta}

\def\p{\partial}

\def\be{\begin{equation}}
\def\ee{\end{equation}}
\def\ba{\begin{align}}
\def\ea{\end{align}}

\def\mg{\sqrt{-g}}

\def\C{\mathcal{C}}
\def\A{\mathcal{A}}

\title{Topological interpretation of extremal and Davies-type phase transitions of black holes}

\author{
Krishnakanta Bhattacharya$^a$\footnote{krish.phys@gmail.com}, \ Kazuharu Bamba$^a$\footnote{bamba@sss.fukushima-u.ac.jp} and Douglas Singleton$^{b}$\footnote{dougs@mail.fresnostate.edu}\\
$^a$Faculty of Symbiotic Systems Science, Fukushima University, Fukushima 960-1296, Japan;\\
$^b$Department of Physics, California State University Fresno, Fresno, CA 93740-8031, USA~.}

\date{\today}
\begin{document}
\maketitle

\begin{abstract}
Topological arguments are currently being used as a novel scheme to discern the properties of black holes while ignoring their detailed structure and specific field equations. Among various avenues of black hole physics, where this novel approach is being utilized, the phase transition in black hole thermodynamics lies at the forefront. There are several types of phase transition in black holes; such as the van der Waals type phase transition, Davies-type phase transition, extremal phase transition, and Hawking-Page (HP) transition.  So far, the topological interpretation, where the critical point has been identified with the non-zero topological charge, has been obtained only for the van der Waals type phase transition and HP transition in different spacetimes. To complete the picture, here we provide the same interpretation for two other phase transitions: Davies-type phase transition and extremal phase transition. The entire analysis is general and is valid for any spacetime where these types of phase transitions are observed. More importantly, our analysis suggests that amid the apparent differences in these phase transitions, they share the same topological characteristics, \textit{i.e.} non-zero topological charge arising from different thermodynamic potentials in different types of phase transition.

\end{abstract}
\section{Introduction}

The initial works of Bekenstein \cite{Bekenstein:1973ur} and Hawking \cite{Hawking:1974sw} first laid the foundation for understanding black holes as thermodynamic objects as it was found that the black holes can be endowed with thermodynamic parameters like entropy and temperature. Furthermore, it was shown that one can obtain the laws of black hole mechanics, which exactly resemble the thermodynamic laws of a conventional thermodynamic system \cite{Bardeen:1973gs}. These initial results were stepping stones to identifying black holes as thermodynamic objects. Even though the microstates that are responsible for black hole thermodynamics remain elusive, we are confident about the thermodynamic connection of black holes.

Phase transitions are one of the key aspects of black hole thermodynamics, which has been explored since the seminal work by Davies \cite{Davies:1989ey}. He identified the critical points as ones with a diverging heat capacity (followed later by \cite{Lousto:1993yr, Lousto:1994cz, Lousto:1994jd, Muniain:1995ih}). Subsequently, in AdS spacetime, another type of phase transition was observed by Hawking and Page \cite{Hawking:1982dh}, known as the Hawking-Page phase transition. Later it was found that the transition of black holes from being non-extermal to extremal marks another type of phase transition, known as the extremal phase transition \cite{curir1,curir2,Pavon:1988in,Pavon:1991kh,Kaburakigrg,Cai:1996df,Cai:1998ep,Wei:2009zzf,Bhattacharya:2019awq}. More recently, in AdS spacetime, considering the cosmological constant as a variable (and identifying the cosmological constant as thermodynamic pressure), a van der Waals-type phase transition is observed in black holes where the mass plays the role of enthalpy \cite{Kastor:2009wy,Dolan:2010ha,Dolan:2011xt,Dolan:2011jm,Dolan:2012jh,Kubiznak:2012wp,Kubiznak:2016qmn,Bhattacharya:2017nru,Bhattacharya:2021lgk}.

Over the years the phase transition of black holes has been studied quite extensively. Earlier, the thermogeometrical interpretation of phase transitions was provided \cite{Cai:1998ep,Quevedo:2008ry,Akbar:2011qw,Hendi:2015cka,Sarkar:2006tg,Hendi:2015xya} (also see the unified approaches \cite{Banerjee:2016nse,Bhattacharya:2017hfj,Bhattacharya:2019qxe}), where the critical points were identified as ones that are endowed with the diverging Ricci-scalar in the thermodynamic phase space. This (thermogeometrical) interpretation has been found to hold good for all types of phase transitions described above. Recently, the topological interpretation has been provided by researchers, which helps to distinguish the properties of the black holes, without resorting to the dynamical equation of the spacetime or the structural details. This formalism was earlier used in the coordinate space of black holes \cite{Cunha:2017qtt,Cunha:2020azh}, to identify the location of the light rings using null geodesics. It was shown that the location of the light rings correspond to the zero point of a special vector, constructed in the coordinate space of the black hole spacetime. Subsequently, for a black hole in asymptotically flat spacetime with a topologically spherical Killing horizon, it was found that the black holes have at least one light ring outside the horizon for each sense of rotation \cite{Cunha:2020azh}, which was later later generalized to other cases \cite{Wei:2020rbh,Guo:2020qwk,Cunha:2024ajc}. Later, this topological argument was used in different aspects of black hole thermodynamics \cite{Wei:2022dzw,Wei:2021vdx,Yerra:2022alz,Yerra:2022eov,Gogoi:2023xzy,Gogoi:2023qku,Yerra:2022coh,Yerra:2023ocu,Yerra:2023,Barzi:2023msl,Ahmed:2022kyv,Wei:2022mzv,Fan:2022bsq,Wu:2022whe,Fang:2022rsb,Wu:2023xpq,Wu:ejpc2023,Wu:2023sue,Li:2023ppc,Wei:2023bgp,Alipour:2023uzo,Zhang:2023uay,Sadeghi:2023aii,Sageghi-2023,Liu:2023sbf,Bai-2023,edery-2021,edery-2022} (also see the recent papers \cite{Lei:2024wvj,Shahzad:2024ojx,Shahzad:2023cis,Wu:2024rmv}). In the black hole phase transition, it was shown that the critical point can have a topological interpretation in terms of non-zero topological charge \cite{Wei:2021vdx,Yerra:2022alz,Yerra:2022eov,Gogoi:2023xzy,Gogoi:2023qku,Yerra:2022coh,Yerra:2023ocu,Barzi:2023msl}. So far, this interpretation has been found to hold for the van der Waals type phase transition \cite{Wei:2021vdx,Yerra:2022alz,Yerra:2022eov,Gogoi:2023xzy,Gogoi:2023qku} and the Hawking-Page phase transition \cite{Yerra:2022coh,Yerra:2023ocu,Barzi:2023msl}. It was first obtained for the van der Waals type phase transition \cite{Wei:2021vdx}, where it was shown that a topological interpretation can be provided using Duan's topological current $\phi$-mapping theory \cite{Duan:1979ucg,duan1}. The same analysis has been extended later to the Hawking-Page transition.  However, the topological interpretation is still missing for the other two types of phase transitions in black holes, namely the Davies type phase transition and the extremal phase transition. Therefore, it becomes an obvious question whether the same interpretation exists for these other two types of phase transitions. In this paper, we will explore this issue, \textit{i.e.} we obtain a topological interpretation for the Davies phase transition and for the extremal phase transition. The entire analysis is general as it is valid for any spacetime which is endowed with these types of phase transitions.

The paper is organized as follows: In section \ref{secgentopo}, we describe Duan's topological current theory, adapted to the thermodynamic phase space. In section \ref{secdavies}, we discuss the topological interpretation of Davies-type phase transition using the topological current theory discussed in section \ref{secgentopo}. In section \ref{secextremal} we provide the same interpretation for the extremal phase transition of black holes. We provide the conclusions of our analysis in section \ref{secconcl}. Throughout our paper, we have adopted the geometrised unit and have set $k_B= \hbar=G=c=1$.
 
\section{Adaptation of Duan’s topological current theory} \label{secgentopo}
The formalism that we follow is based on Duan's $\phi$-mapping topological current theory \cite{Duan:1979ucg,duan1}, originally defined in the coordinate space, where the coordinates are defined as $x^{\mu}=\{\tau, x^a\}$. Thus, in the following discussion, greek indices ($\mu$,$\nu$, $\rho$ \textit{etc.}) run 0,1,2 and the roman indices run 1,2 \footnote{Note that Duan's formalism \cite{Duan:1979ucg,duan1} is defined in (1+3) coordinate space, which we adapt to (1+2) space as our goal is to use this 
 in the thermodynamic phase space in the following sections \textit{i.e.} in Section \ref{secdavies} and Section \ref{secextremal}, where $x^a$ will correspond to the thermodynamic and associate variables.}. At the onset, one defines a potential $\boldsymbol{\Phi} (x^a)$, which is known as
Duan’s potential. Thereafter, the vector field ($\phi$) is defined as $\phi=\{\phi^a\}=\{\phi^1,\phi^2\}$ where 
\begin{align}
    \phi^a=\p_a\boldsymbol{\Phi}~.
\end{align}
The normalized vector $n^a$ can be defined as
\begin{align}
    n^a=\frac{\phi^a}{||\phi||}~,\label{n}
\end{align}
which has the properties 
\begin{align}
    n^an^a=1~~~~~~~~~~~~~~~~~~\textrm{and}~~~~~~~~~~~~~~~~~~ n^a\p_{\nu}n^a=0~. \label{nproperties}
\end{align}
Using the above-normalized vector $n^a$, one can define an anti-symmetric superpotential $V^{\mu\nu}$ as 
\begin{align}
    V^{\mu\nu}=\frac{1}{2\pi}\epsilon^{\mu\nu\rho}\epsilon_{ab}n^a\p_{\rho}n^b~.\label{Vmunu}
\end{align}
 Employing the above anti-symmetric superpotential, one can obtain a topological current $j^\mu$ as 
\begin{align}
    j^{\mu}=\p_{\nu} V^{\mu\nu}=\frac{1}{2\pi}\epsilon^{\mu\nu\rho}\epsilon_{ab}\p_{\nu}n^a\p_{\rho}n^b \label{jmu}
\end{align}
Due to the anti-symmetric property of $V^{\mu\nu}$, it is straightforward to find that the current $j^\mu$ is conserved, \textit{i.e.}
\begin{align}
    \p_{\mu}j^\mu=0~.\label{conservation}
\end{align}
 The zeroth component of $j^{\mu}$ can be thought of as the charge density. Therefore, the total total charge $Q$ within a region $\C$ can be obtained as
 \begin{align}
     Q=\int_{\C}j^0d^2x~. \label{Q}
 \end{align}
The important properties of the charge $Q$ can be unveiled with the following analysis. Substituting \eqref{n} in \eqref{jmu}, one can obtain
\begin{align}
    j^{\mu}=\frac{1}{2\pi}\epsilon^{\mu\nu\rho}\epsilon_{ab}\frac{\p}{\p\phi^c}\Big(\frac{\phi^a}{||\phi^2||}\Big)\p_{\nu}\phi^c\p_{\rho}\phi^b~. \label{jmu1}
\end{align}
One can further obtain 
\begin{align}
    \frac{\phi^a}{||\phi^2||}=\frac{\p\ln ||\phi||}{\p\phi^a}~. \label{phia}
\end{align}
Substituting \eqref{phia} on \eqref{jmu1}, one can obtain the expression of the conserved current as
\begin{align}
    j^{\mu}=\frac{1}{2\pi}\epsilon^{\mu\nu\rho}\epsilon_{ab}\Big(\frac{\p}{\p\phi^c}\frac{\p}{\p\phi^a}\ln ||\phi||\Big)\p_{\nu}\phi^c\p_{\rho}\phi^b~.
\end{align}
The above expression of topological current can be simplified by introducing the Jacobi tensor, which is given as
\begin{align}
    \epsilon^{cb} J^{\mu}\Big(\frac{\phi}{x}\Big)=\epsilon^{\mu\nu\rho}\p_{\nu}\phi^c\p_{\rho}\phi^b~,\label{jacobi}
\end{align}
where $J(\frac{\phi}{x}) \equiv \frac{\partial(\phi^1,\phi^2)}{\partial(x^1,x^2)}$ is the Jacobian.  
Thus, the expression of $j^{\mu}$ can be obtained in terms of Jacobi tensor as
\begin{align}
    j^{\mu}=\frac{1}{2\pi}\Big(\Delta_{\phi}\ln ||\phi||\Big)J^{\mu}\Big(\frac{\phi}{x}\Big)~,
\end{align}
where $\Delta_{\phi}=\frac{\p}{\p\phi^a}\frac{\p}{\p\phi^a}$~. Using the two-dimensional Laplacian Green function of $\phi$ space ($\Delta_{\phi}\ln ||\phi||=2\pi\delta^2(\phi)$), one can finally obtain the expression of the current $j^{\mu}$ as
\begin{align}
    j^{\mu}=\delta^2(\phi)J^{\mu}\Big(\frac{\phi}{x}\Big)~,
\end{align}
and the total charge inside $\C$ is given as
\begin{align}
    Q=\int_{\C}\delta^2(\phi)J^{0}\Big(\frac{\phi}{x}\Big)d^2x~. \label{Qfinal}
\end{align}
Thus, the above expressions suggest that the topological current $j^{\mu}$ is non-vanishing only at $\phi=0$ (\textit{i.e.} on the zeros of $\phi$) and the topological charge $Q$ is non-vanishing \textit{iff } $\C$ encloses at lease one zero point of $\phi$ \footnote{Note that for those zeros of $\phi$, which are regular, it can be argued by implicit function theorem \cite{goursat} that $J^{0}\Big(\frac{\phi}{x}\Big)\neq 0$. Our analysis is valid for the regular zeros. However, if at the zero point $J^{0}\Big(\frac{\phi}{x}\Big)= 0$, which might happen if the zeros of $\phi$ include some branch points, one has to apply the bifurcation theory of topological current. For more details on it, see \cite{Duan:1998kw}.}. If $\C$ encloses no zero point, the corresponding charge vanishes. Furthermore, if two boundary curves $\C_1$ and $\C_2$ enclose the same zero points, the corresponding topological charge will be the same. 

Furthermore, the above topological charge can be expressed in terms of the winding number as per the following analysis. Suppose there are $N$ isolated zero points of $\phi$, and the $i$-th zero is located at $x=z_i$~. Then from the theory of delta function \cite{A.S.Schwarz}, one can obtain 
\begin{align}
    \d^2(\phi)=\sum_{i=1}^N\frac{1}{\Big|J^0\Big(\frac{\phi}{x}\Big)\Big|_{x=z_i}}\d^2(x-z_i)~.
\end{align}
Now, we consider that when $x$ goes around the zero point $z_i$ once, the number of loops that $\phi^a$ makes in the $\phi$ vector space is given by the positive integer number $\beta_i$ (which is known as the Hopf index in topology). Then one can further obtain 
\begin{align}
    \delta^2(\phi)J^{0}=\sum_{i=1}^N \beta_i\eta_i\d^2(x-z_i)~,
\end{align}
On the right hand side $\beta_i$ arises due to accounting the fact that when $x$ goes around the $i^{th}$ zero once, the number of loop made in the $\phi$ space is given by the Hopf index $\beta_i$ \cite{duan1}.
where the Brower degree $\eta_i$ is defined as
\begin{align}
    \eta_i=\left[\frac{J^0\Big(\frac{\phi}{x}\Big)}{\Big|J^0\Big(\frac{\phi}{x}\Big)\Big|}\right]_{x=z_i}=\textrm{sign}\Big(J^0\Big(\frac{\phi}{x}\Big)_{z_i}\Big)=\pm 1~.
\end{align}
Finally, the total charge inside $\C$ can be obtained as
\begin{align}
    Q=\sum_{i=1}^N \beta_i\eta_i=\sum_{i=1}^N w_i~,
\end{align}
where $w_i=\beta_i\eta_i$ is known as the winding number of $i$-th zero point contained in $\C$~. If one further considers that the $i$-th zero point is enclosed by a smooth and positively oriented closed curve $\C_i$ and there are no other zero points inside $\C_i$, one can obtain the winding number corresponding to the $i$-th zero point given as
\begin{align}
    w_i=\frac{1}{2\pi}\oint_{\C_i}d\Omega~,
\end{align}
where $\Omega$ is known as the deflection angle of the vector field along the given contour and it is given by $\Omega=\arctan(\phi^2/\phi^1)$. In terms of the unit vectors $n^1$ and $n^2$, $d\Omega$ can be obtained as $d\Omega=n^1dn^2-n^2dn^1=\epsilon_{ab}n^adn^b$ (the expression which will be used later).

With the above mathematical preliminaries, we can now move on to the main analysis \textit{i.e.} topological interpretation of critical points of black hole phase transition. Our major goal is to find a suitable thermodynamic potential $\boldsymbol{\Phi}$ that yields vectors $\phi^a=\p_a\boldsymbol{\Phi}$ such that $\phi^a$ vanishes on the critical point under consideration.

\section{Davies-type phase transition} \label{secdavies}
As it is now well-known, black holes behave like thermodynamic objects and are endowed with the first law of the form
\begin{align}
    TdS=dM-\sum_i Y_idX_i~, \label{1stlaw}
\end{align}
where $X_i$ are the generalized displacements like angular momentum ($J$), electric charge ($Q$) and $Y_i$ corresponds to the generalized forces like angular velocity ($\Omega$), electric potential ($\Phi_E$) \textit{etc.}

For Davies-type phase transition \cite{Davies:1989ey,Lousto:1993yr, Lousto:1994cz, Lousto:1994jd, Muniain:1995ih}, one looks for the divergence of the specific heat, which is defined as
\begin{align}
    C_{X_i}=T\Big(\frac{\p S}{\p T}\Big)_{X_i}~.
\end{align}
Now, using 
\begin{align}
    \frac{\p}{\p S}\Big(\frac{1}{T}\Big)=-\frac{1}{T^2}\Big(\frac{\p T}{\p S}\Big)_{X_i}~,
\end{align}
one can obtain 
\begin{align}
    C_{X_i}=-\frac{1}{T\frac{\p}{\p S}\Big(\frac{1}{T}\Big)_{X_i}}
\end{align}
Since $T\neq 0$, at the critical point, one obtains 
\begin{align}
    \frac{\p}{\p S}\Big(\frac{1}{T}\Big)_{X_i}=0~. \label{CRICCON}
\end{align}
In the following, we provide the topological description of Davies-type phase transition adopting Duan's formalism mentioned in the previous section. As we have mentioned earlier, Duan's original formalism is defined in the coordinate space, which we shall use in the thermodynamic space of the black hole. More importantly, here we use the concept of topological charge, which is originally defined in the coordinate space in the thermodynamic phase space. Taking a cue from Eq. \eqref{CRICCON}, we define the thermodynamic potential $\boldsymbol{\Phi}$ as
\begin{align}
    \boldsymbol{\Phi}(S,\theta)=\frac{1}{T(S)_{X_i}\sin\theta}~.
\end{align}
Here the additional $1/\sin\theta$ factor has been introduced in the thermodynamic potential (adopting the idea from \cite{Wei:2021vdx}) which simplifies the study of topology. The vector fields can be obtained as 
\begin{align}
    \phi^S=\p_S \boldsymbol{\Phi}(S,\theta)=\frac{1}{\sin\theta}\frac{\p}{\p S}\Big(\frac{1}{T}\Big)_{X_i}~,
    \no 
    \\
    \phi^{\theta}=\p_{\theta}\boldsymbol{\Phi}(S,\theta)=-\frac{\csc\theta\cot\theta}{T(S)_{X_i}}
\end{align}
Now we can realize the significance of the auxiliary term $1/\sin\theta$. It serves two major purposes: (i) The vector $\phi$ is perpendicular to the horizontal lines $\theta=0$ and $\theta=\pi$, which can be considered as the natural boundary of the parameter space. (ii) More importantly, this auxiliary term helps to identify the critical point on the $S-\theta$ plane. Since $\phi^{\theta}$ vanishes at $\theta=\pi/2$~,the zero point of $\phi$ (where both $\phi^S$ as well as $\phi^{\theta}$ should vanish) would lie on $\theta=\frac{\pi}{2}$ line. In other words, the zero point of $\phi$ in the $S-\theta$ plane (which can easily be identified on $\theta=\frac{\pi}{2}$ line due to the presence of the auxiliary term), will correspond to the critical point where $\phi^{\theta}$ vanishes due to $\theta=\frac{\pi}{2}$ and $\phi^S$ vanishes due to the relation \eqref{CRICCON} on the critical point.  Let us also mention that although we have adopted the auxiliary term as $1/\sin\theta$ following \cite{Wei:2021vdx}, it can also be taken as $1/\cos\theta$ as well. In that case, the natural $\theta$ limit would be $-\pi/2$ to $\pi/2$ and the critical point will be located at $\theta=0$ line on the $S-\theta$ plane. The zero point of $\phi$ on  $\theta=\pi/2$ line corresponds to the critical point where $\phi^S$ vanishes due to the criticality condition and $\phi^{\theta}$ vanishes as $\theta=\pi/2$. Thus, the critical point can be attributed to a non-zero topological charge using the formalism mentioned in the previous section.

Thus, from the above general analysis, we can obtain the topological interpretation of the Davies-type phase transition, which will be applicable to any spacetime that has a Davies-type critical point. On the critical point, one can obtain a zero point as per our analysis and one can attribute a non-zero topological charge.

Let us now realize this with a particular example. Let us consider the following action \cite{Jardim:2012se}
\begin{align}
    \A=\int d^4x\mg [\frac{1}{16\pi}\Big(R+2\Lambda)+ \frac{1}{4}\eta F^{\mu\nu}F_{\mu\nu}]~,
\end{align}
where the first term is the Einstein-Hilbert term, the second term characterizes the dS ($\Lambda> 0$), AdS ($\Lambda< 0$) or asymptotically flat ($\Lambda=0$) spacetime depending upon the value of $\Lambda$ and the third term is the coupling of the Maxwell field (for $\eta=1$) or phantom field (for $\eta=-1$). The metric solution of the above action is given as \cite{Jardim:2012se}
\begin{align}
    ds^2=-f(r) dt^2+\frac{dr^2}{f(r)}+r^2d\Omega_2^2~,
\end{align}
where 
\begin{align}
    f(r)=1-\frac{2M}{r}-\frac{\Lambda}{3}r^2+\eta \frac{Q^2}{r^2}~. \label{phantomsol}
\end{align}
The horizons of the solution can be obtained by solving the condition $f(r)=0$, which yields two roots $r_+$ and $r_-$ and the outer horizon can be identified as the event horizon \cite{Jardim:2012se}. The mass of the black hole can be expressed in terms of other thermodynamic parameters as
\begin{align}
    M=\frac{1}{2}\Big(\frac{S}{\pi}\Big)^{\frac{3}{2}}\Big(\frac{\pi}{S}-\frac{\Lambda}{3}+\frac{\eta\pi^2Q^2}{S^2}\Big)~,
\end{align}
where $S=\pi r_+^2$ is the Bekenstein-Hawking entropy. The Hawking temperature can be obtained as 
\begin{align}
    T=\Big(\frac{\p M}{\p S}\Big)_Q=\frac{-\pi ^2 \eta  Q^2-\Lambda  S^2+\pi  S}{4 \pi ^{3/2} S^{3/2}}~.
\end{align}
The above expression of the Hawking temperature can also be obtained from the surface gravity ($\kappa$) as $T=\kappa/2\pi$, where $\kappa=f'(r_+)/2$~. The specific heat ($C_Q$) can be obtained as
\begin{align}
    C_Q=T\Big(\frac{\p S}{\p T}\Big)_Q=\frac{2 S \left(\pi ^2 \eta  Q^2+\Lambda  S^2-\pi  S\right)}{-3 \pi ^2 \eta  Q^2+\Lambda  S^2+\pi  S}~.\label{cq}
\end{align}
From the above expression, one can obtain the divergence points  of specific heat (or the Davies points) when the denominator of \eqref{cq} vanishes and one obtains the expression of entropy at the Davies points as \footnote{Since it is a quadratic equation, it has two roots. The other value of entropy can be discarded as it is negative in value.}
\begin{align}
    S_D=\frac{\pi}{2\Lambda}\Big(-1+\sqrt{1+12\eta\Lambda Q^2}\Big)~.
\end{align}
For simplicity, we fix the value of $Q$, $\Lambda$, $\eta$ as $Q=\eta=\Lambda=1$. In that case, the value of entropy on the Davies point can be obtained as $S_D=4.0927$. 

\begin{figure}
\centering
\includegraphics[width=0.54\textwidth]{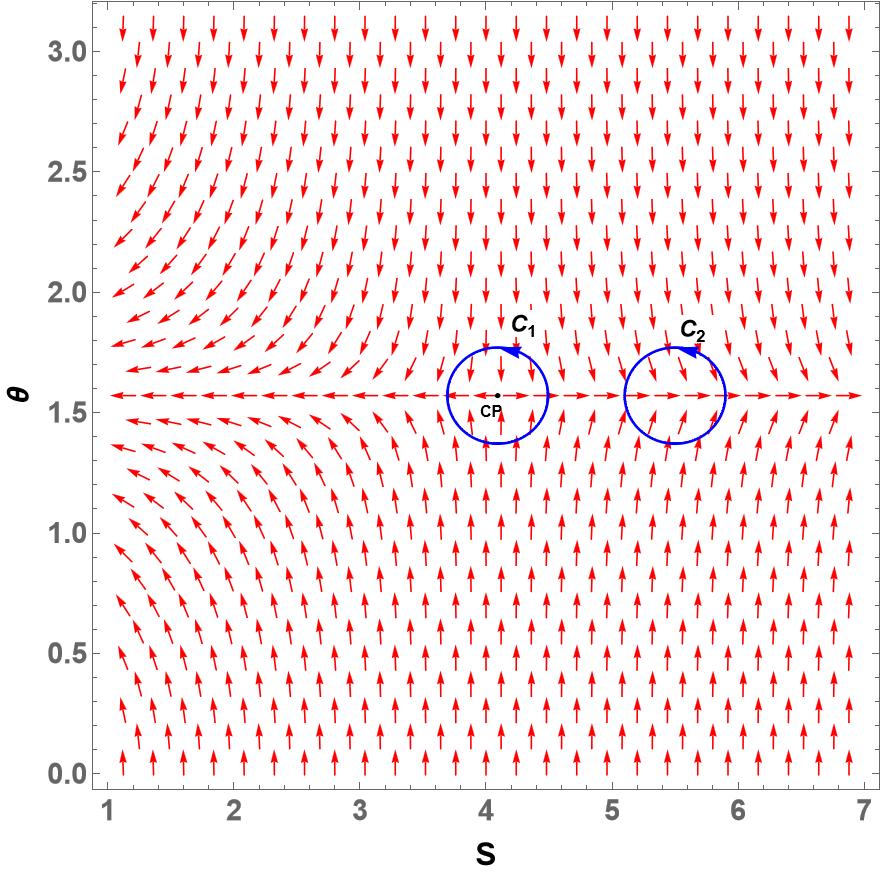}
\caption{The red arrow represents the vector $n^a$ in the $S-\theta$ plane. The Davies critical point has been denoted by ''CP'' (black dot), which is located at $S=S_D,\theta=\frac{\pi}{2}$. The two contours (blue circles) $C_1$ and $C_2$ are the two closed loops where $C_1$ encircles the critical point CP and $C_2$ does not.}
\label{pltdavies}
\end{figure}

 Let us now find the topological interpretation of the Davies-type phase transition. As per our formalism, the thermodynamic potential can be defined as
 \begin{align}
     \boldsymbol{\Phi}(S,\theta)=\frac{1}{\sin\theta}\frac{1}{T}\Big|_{Q=\eta=\Lambda=1}=\frac{4 \pi ^{3/2} S^{3/2}}{\sin\theta(-S^2+\pi  S-\pi ^2)}~.
 \end{align}
 The components of the vector field $\phi$ can be obtained as
 \begin{align}
     \phi^S=-\frac{2 \pi ^{3/2} \sqrt{S} \left(-S^2-\pi  S+3 \pi ^2\right) \csc\theta}{\left(S^2-\pi  S+\pi ^2\right)^2}~,
     \no 
     \\
     \phi^{\theta}=\frac{4 \pi ^{3/2} S^{3/2} \cot\theta \csc\theta}{S^2-\pi  S+\pi ^2}~.~~~~~~~~~~~~~~~~
 \end{align}
 The normalized vector can be obtained as $n^a=(\frac{\phi^S}{||\phi||}, \frac{ \phi^{\theta}}{||\phi||})$, which is represented in figure \ref{pltdavies}, which is a pictorial representation of the unit vector $n$ in the $S-\theta$ plane. 
 From the figure \ref{pltdavies}, we find that the zero point (which can be marked as the Davies-type critical point) is located at $S=S_D$ and $\theta=\frac{\pi}{2}$~. Since the winding number or the topological charge is independent of the loops enclosing the zero point, we consider two circular loops $C_1$ and $C_2$ where $C_1$ encloses the critical point and $C_2$ does not. The two contours are parameterised by $\psi$ and are defined as follows:
 \begin{align}
     S=a \cos\psi+S_0~,
     \no 
     \\
     \theta=a\sin\psi+\frac{\pi}{2}~, \label{sntheta}
 \end{align}
 which represents circle of radius $a$ and centred at ($S_0, \frac{\pi}{2}$). For $C_1$, we choose $a=0.4$ and $S_0=S_D$, whereas for $C_2$, we choose $a=0.4$ and $S_0=5.5$. The deflection angle can be measured using the definition provided in the previous section, which can be obtained as 
 \begin{align}
     \Omega(\psi)=\int_0^{\psi}\epsilon_{ab} n^a\p_{\psi}n^bd\psi~,
 \end{align}
and the topological charge can be obtained as
\begin{align}
    Q=\frac{\Omega(2\pi)}{2\pi}~. \label{Qomega}
\end{align}
For our case, we find that for $C_1$, the topological charge can be obtained as $Q=-1$, and for $C_2$, the topological charge can be obtained as $Q=0$. Thus, using our formalism, the Davies-type critical points can be given a topological interpretation and the Davies points can be associated with non-zero topological charge. Note that Choosing $S_0 = 5.5$ is just for illustrative purposes. Any contour that does not include the Davies point will have a zero topological charge. Furthermore, the positivity or the negativity of charges will also depend on the choice of contour as well. In our case, we have considered the contour to be counter-clockwise. If the contour encloses the zero-point clockwise, we would have obtained $Q=+1$.

\section{The extremal phase transition} \label{secextremal}
In the extremal phase transition, the black hole undergoes a transition from being non-extremal to extremal one and the critical point is located at the extremal limit. In the extremal limit, the surface gravity of the black hole (or the black hole temperature) vanishes. In this case, the desired thermodynamic potential can be identified in terms of the thermodynamic mass of the black hole as 
\begin{align}
    \boldsymbol{\Phi}(S,\theta)=\frac{M(S)_{X_i}}{\sin\theta}~.
\end{align}
Again the auxiliary factor $1/\sin\theta$ has been included for the reasons mentioned earlier. The vector fields can be obtained as
\begin{align}
        \phi^S=\p_S \boldsymbol{\Phi}(S,\theta)=\frac{1}{\sin\theta}\Big(\frac{\p M}{\p S}\Big)_{X_i}~,
    \no 
    \\
    \phi^{\theta}=\p_{\theta}\boldsymbol{\Phi}(S,\theta)=-M\csc\theta\cot\theta
\end{align}
Again, we find that the zero point of $\phi$ on $\theta=\frac{\pi}{2}$ line is located on the critical point where $T=(\p M/\p S)_{X_i}$ vanishes. Therefore, applying the discussion presented in the section \ref{secgentopo}, this general analysis suggests that the extremal critical point can be attributed to a non-zero topological charge.

The above analysis will be valid for any spacetime which has an extremal critical point, where the Hawking temperature $T$ vanishes. However, we shall again realize it for a given spacetime. Let us consider the BTZ black hole \cite{Banados:1992wn,Banados:1992gq}, which is the solution of Einstein's gravity in ($1+2$) dimensional spacetime. The metric is given as
\begin{align}
    ds^2=-N^2(r)dt^2+\frac{dr^2}{N^2(r)}+r^2\Big(N^{(\rho)}dt+d\rho\Big)^2~.
\end{align}
where
\begin{align}
    N^2(r)=-M+\frac{r^2}{l^2}+\frac{J^2}{4r^2}, ~~~~~~N^{(\rho)}(r)=-\frac{J}{2r^2}~.
\end{align}
Here $M$ and $J$ are the mass and angular momentum of the black hole and $l$ is related to the cosmological constant. The horizons can be obtained from the condition that the lapse function ($N$) which is given as
\begin{align}
    r_{\pm}=\Big[\frac{Ml^2}{2}(1\pm\Delta)\Big]^{\frac{1}{2}}~, ~~~~~~~~\Delta=\Big[1-\left(\frac{J}{Ml}\right)^2\Big]^{\frac{1}{2}}~. \label{rpmdelta}
\end{align}
The extremal condition is given by $\Delta\longrightarrow 0$, in which case the outer horizon (or the event horizon, denoted by $r_+$) and the inner horizon (denoted by $r_-$) coincide. The Hawking temperature of the black hole is given as
\begin{align}
    T=\frac{\kappa}{2\pi}=\frac{r_+^2-r_-^2}{2\pi r_+l^2}~,
\end{align}
which as mentioned above vanishes in the extremal limit when $r_+ = r_-$. The expression of entropy is given by \cite{Banados:1992wn,Banados:1992gq}
\begin{align}
    S=4\pi r_+=4\pi \Big[\frac{Ml^2}{2}(1+\Delta)\Big]^{\frac{1}{2}}~.
\end{align}
The expression for the mass ($M$) as a function of entropy can be expressed as
\begin{align}
    M=\frac{64 \pi ^4 J^2 l^2+S^4}{16 \pi ^2 l^2 S^2}~.
\end{align}
The thermodynamic potential $\boldsymbol{\Phi}$ can be identified as 
\begin{align}
    \boldsymbol{\Phi}(S,\theta)=\frac{M(S)\Big|_{J=1, l=1}}{\sin\theta}=\frac{S^4+64 \pi ^4 }{16 \pi ^2 S^2\sin\theta}~.
\end{align}
Note that here we have set $J=1=l$ for our convenience. From \eqref{rpmdelta}, we find that at the extremal limit, the mass is given as $M_{ext}=J/l$. For $J=1=l$, we find that at the extremal limit, the value of mass and entropy is given as $M_{ext}=1$ and $S_{ext}=2 \sqrt{2} \pi=8.88577$. 

\begin{figure}
\centering
\includegraphics[width=0.74\textwidth]{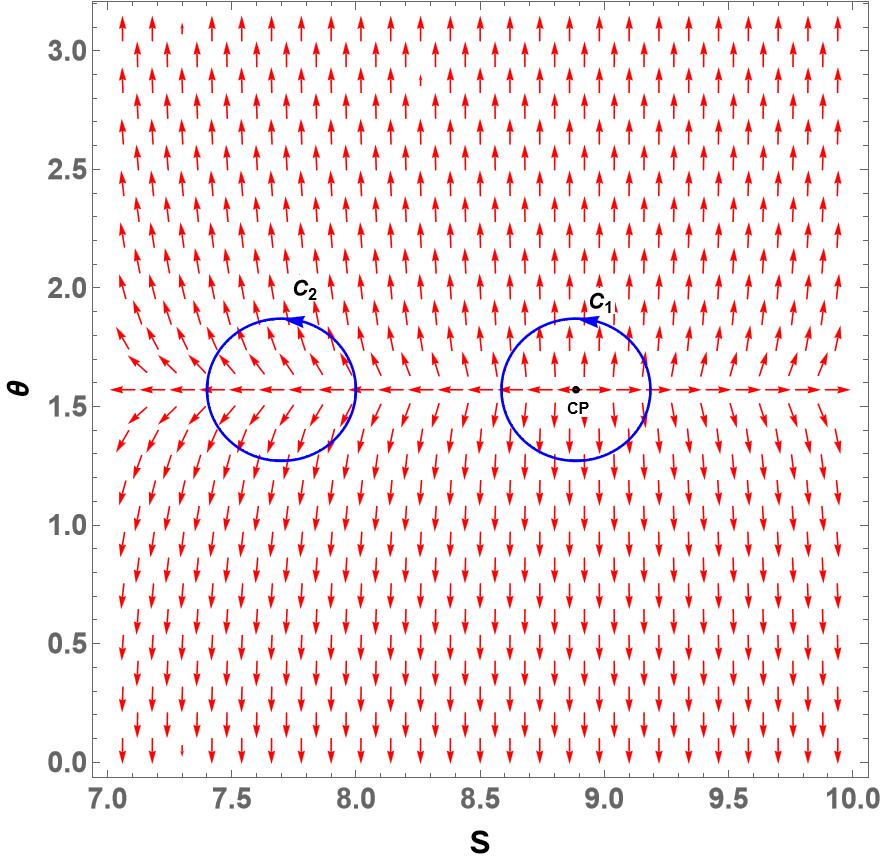}
\caption{The red arrows represent the vector $n^a$ in the $S-\theta$ plane. The extremal critical point has been denoted by ``CP" (black dot), which is located at $S=S_{ext},\theta=\frac{\pi}{2}$. The two contours (blue circles) $C_1$ and $C_2$ are the two closed loops where $C_1$ encircles the critical point CP and $C_2$ does not.}
\label{pltext}
\end{figure}

The components of the vector field $\phi$ can be obtained as
\begin{align}
     \phi^S=\left(\frac{S}{8 \pi ^2}-\frac{8 \pi ^2}{S^3}\right) \csc\theta~, ~~~~~~
     \no 
     \\
     \phi^{\theta}=-\frac{\left(S^4+64 \pi ^4\right)}{16 \pi ^2 S^2}\cot\theta\csc\theta~.
\end{align}
Again, the normal fields can be obtained as $n^a=(\frac{\phi^S}{||\phi||}, \frac{ \phi^{\theta}}{||\phi||})$, which is again represented in figure \ref{pltext}, which is a pictorial representation of the unit vector $n$ in the $S-\theta$ plane.

From the figure \ref{pltext}, one can find that the critical point is located at $S=S_{ext}$ and $\theta=\pi/2$~.

To obtain the topological charge, again we choose two loops $C_1$ and $C_2$, which are defined by the parametric relation \eqref{sntheta}. For $C_1$, we choose $a$ and $S_0$ as $a=0.3$ and $S_0=S_{ext}$, which implies that $C_1$ encloses the critical point. For $C_2$, we choose $a$ and $S_0$ as $a=0.3$ and $S_0=7.7$, which implies that $C_2$ does not enclose the critical point. For these loops, we calculate the topological charge (using eq. \ref{Qomega}). It turns out that for $C_2$, the topological charge vanishes but for $C_1$, one can obtain $Q=1$~.

Thus, using our formalism, one can provide the topological interpretation of the extremal phase transition and the extremal critical point can be associated with a non-zero topological charge. Note that for the specific example in Davies type phase transition, we obtained $Q=-1$, whereas for the present example in extremal phase transition, we have obtained $Q=+1$. As we discussed earlier, it depends upon the orientation of the $\phi^a$ vector (as it can be seen, the direction of arrows in figure \ref{pltdavies} and figure \ref{pltext} are different whereas the contours are counter-clockwise in every case). Earlier, employing the generalized off-shell free energy, the solutions of black holes were given topological interpretation as the defects in the thermodynamic parameter space \cite{Wei:2022dzw}. In that case, using free energy as the thermodynamic potential in every case, it was discussed that black holes can be classified into three groups based on their topological charges as $Q=-1,0,+1$~. We find that the same thermodynamic potential cannot provide the topological interpretation for every type of phase transition in black holes (For Davies-type phase transition, inverse-temperature, and for extremal phase transition black hole mass plays the role of the thermodynamic potential in our case). Furthermore, in different cases, we obtain both $Q=-1$ and $Q=+1$. These mark the similarities and differences of our analysis with \cite{Wei:2022dzw}.

\section{Conclusions:} \label{secconcl}

Over the years, the study of phase transitions in black hole thermodynamics has gained a lot of prominence among physicists. It is considered one of the most fascinating parts of the theory of black hole physics.  Different types of phase transitions have been observed in black holes: Davies-type phase transitions, Hawking-Page phase transitions, extremal phase transitions, and van der Waals-type phase transitions. In this paper, we have presented the topological interpretation for Davies-type and extremal phase transition, which was missing in the literature thus far.

 Recently, it was found that the thermodynamic properties of black holes can be deduced when one uses a phenomenological theory of topology in the thermodynamic phase space. At first \cite{Wei:2022dzw}, it was shown that all the black hole solutions can be identified in three groups based on the topological charges defined in the thermodynamic parameter space \textit{i.e.} $Q=-1, 0,$ and $ +1$. In addition, it was shown that the black holes with the same topological nature exhibit similar thermodynamic properties. Later, this topological argument has been extended to explore the phase transitions in black holes. Firstly, it was shown in \cite{Wei:2021vdx} that the critical point in the Van der Waals type phase transition of the black holes can have a topological interpretation using Duan's $\phi$-field formalism in the thermodynamic parameter space. In that case, the critical point can be attributed to a non-zero topological charge. Later, this argument has been extended to the Hawking-Page phase transition \cite{Yerra:2022coh}. However, the topological interpretation is still missing in literature for the other two types of phase transitions \textit{i.e.} the Davies-type phase transition and the extremal phase transition. Thus, inspired by \cite{Wei:2021vdx}, we have adopted Duan's $\phi$-mapping topological current theory to provide the topological interpretation for Davies-type and extremal phase transitions. Thereby, we show that the topological interpretation can indeed be provided for these two types of phase transitions by making judicious choices of the thermodynamic potential $\boldsymbol{\Phi}$ in every case. For the Davies-type phase transition, which is characterized by the divergence of heat capacity, the inverse Hawking temperature ($T^{-1}$) plays the role of the thermodynamic potential to obtain the topological interpretation. On the other hand, for the extremal phase transition, we found that the mass of the black hole plays the role of the thermodynamic potential. In both cases, we have chosen the thermodynamic potential in such a way that the critical point lies on the zero-point of the vector field $\phi^a$, which is obtained from the thermodynamic potential $\boldsymbol{\Phi}$. Despite the entire description being provided from a general standpoint, we also have provided specific examples to support our claim and have shown that the topological argument holds good for the black hole phase transition. In all cases, we find that the critical point can be endowed with non-zero topological charges using the $\phi$-field formalism. More importantly, we show that the recent development of topological interpretation of black hole phase transitions, which has been so far explored for van der Waals and Hawking-Page phase transition, can be extended for other criticalities as well. Thus, the topological interpretation can act as a standard formalism to describe the phase transition in black holes.

Finally, as we have mentioned earlier, the topological argument is provided in order to describe the properties of the black holes without involving the field equation and detail structure. Hence, the formalism mentioned here can be applicable beyond Einstein's GR and to any black holes that show these types of phase transitions. Let us explain it in a bit more detail. The thermodynamic properties of black holes were initially established in general relativity \cite{Bekenstein:1973ur,Hawking:1974sw,Bardeen:1973gs}. However, later it was found that the thermodynamic properties are valid well beyond general relativity \cite{Eling:2006aw,Elizalde:2008pv,Chirco:2009dc,Padmanabhan:2002sha,Padmanabhan:2003gd,Faraoni:2010yi,Bamba:2009gq,Jacobson:1993vj,Iyer:1994ys,Visser:1993nu} (for a recent review on this matter, see \cite{Bhattacharya:2022mnb}). Phase transition, which is a key aspect of black hole thermodynamics, has been found beyond GR as well. For instance, Van der Waals-type phase transition has been observed in higher curvature gravity and scalar-tensor gravity \cite{Chen:2013ce,Ovgun:2017bgx,Hendi:2015kza,Hendi:2015hgg}. Similarly, Davies-type phase transition has been observed beyond general relativity  such as in Horava-Lifshitz gravity \cite{Majhi:2012fz}, Gauss-Bonnet gravity \cite{Du:2019poh}, Lovelock gravity \cite{Ali:2023zgm}, $f(R)$ gravity \cite{EslamPanah:2024tex}, and so on. Furthermore, one can also find extremal black hole solutions beyond GR \cite{Ayuso:2020rmb} and one can expect to find extremal phase transition beyond general relativity as well. We emphasize that our analysis will continue to be valid for all those cases as our topological argument has been formulated in the thermodynamic parameter space and does not involve any gravitational equation.

Since our topological argument is formulated in the thermodynamic parameter space of a black hole (BH), we admit that the observational connection with our analysis is not direct and we need to investigate much deeper to comment on the potential connection with black hole shadow, accretion disc \textit{etc.} However, our analysis distinguishes certain black holes from others, which can have significance from the observational viewpoint. For example, it distinguishes extremal black holes (non-zero topological charge) from the non-extremal ones. Therefore, the extremal black holes can be identified from our analysis. Furthermore, we discussed that for Davies-type phase transition, heat capacity diverges for certain black holes at a specific horizon radius. A closer investigation (see \cite{TranNHung:2024pig} for example) shows that the larger BHs show positive heat capacity and the smaller BHs show negative heat capacity and at the critical limit, heat capacity diverges. Therefore, smaller BHs are relatively unstable. From our topological argument, one can obtain the stable limit (non-zero topological charge) of different black holes. To obtain further connection with other observable features, we need to investigate further, which we keep as a potential future work.

\section*{Acknowledgement} This work is supported by the JSPS KAKENHI Grant (Number: 23KF0008) and the research of K. Bamba is supported in part by the JSPS KAKENHI Grant (Number: JP21K03547).


\begin{thebibliography}{99}


%\cite{Bekenstein:1973ur}
\bibitem{Bekenstein:1973ur} 
  J.~D.~Bekenstein,
  ``Black holes and entropy,''
  Phys.\ Rev.\ D {\bf 7}, 2333 (1973).
  %doi:10.1103/PhysRevD.7.2333  
  
  %\cite{Hawking:1974sw}
\bibitem{Hawking:1974sw} 
  S.~W.~Hawking,
  ``Particle Creation by Black Holes,''
  Commun.\ Math.\ Phys.\  {\bf 43}, 199 (1975)
  Erratum: [Commun.\ Math.\ Phys.\  {\bf 46}, 206 (1976)].
  %doi:10.1007/BF02345020
  %%CITATION = doi:10.1007/BF02345020;%%
  
  %\cite{Bardeen:1973gs}
\bibitem{Bardeen:1973gs} 
  J.~M.~Bardeen, B.~Carter and S.~W.~Hawking,
  ``The Four laws of black hole mechanics,''
  Commun.\ Math.\ Phys.\  {\bf 31}, 161 (1973).
  %doi:10.1007/BF01645742
  %%CITATION = doi:10.1007/BF01645742;%%
 
   
  %\cite{Davies:1989ey}
\bibitem{Davies:1989ey} 
  P.~C.~W.~Davies,
  ``Thermodynamic Phase Transitions of {Kerr-Newman} Black Holes in De Sitter Space,''
  Class.\ Quant.\ Grav.\  {\bf 6}, 1909 (1989).
  %doi:10.1088/0264-9381/6/12/018
  %%CITATION = doi:10.1088/0264-9381/6/12/018;%%

  %\cite{Lousto:1993yr}
\bibitem{Lousto:1993yr} 
  C.~O.~Lousto,
  ``The Fourth law of black hole thermodynamics,''
  Nucl.\ Phys.\ B {\bf 410}, 155 (1993)
  Erratum: [Nucl.\ Phys.\ B {\bf 449}, 433 (1995)]
  %doi:10.1016/0550-3213(93)90577-C, 10.1016/0550-3213(95)00322-J
  [gr-qc/9306014].
  
  %\cite{Lousto:1994cz}
\bibitem{Lousto:1994cz} 
  C.~O.~Lousto,
  ``Effective two-dimensional description from critical phenomena in black holes,''
  Gen.\ Rel.\ Grav.\  {\bf 27}, 121 (1995).
  
  %\cite{Lousto:1994jd}
\bibitem{Lousto:1994jd} 
  C.~O.~Lousto,
  ``The Emergence of an effective two-dimensional quantum description from the study of critical phenomena in black holes,''
  Phys.\ Rev.\ D {\bf 51}, 1733 (1995)
  %doi:10.1103/PhysRevD.51.1733
  [gr-qc/9405048].
  
  %\cite{Muniain:1995ih}
\bibitem{Muniain:1995ih} 
  J.~P.~Muniain and D.~D.~Piriz,
  ``Critical behavior of dimensionally continued black holes,''
  Phys.\ Rev.\ D {\bf 53}, 816 (1996)
  %doi:10.1103/PhysRevD.53.816
  [gr-qc/9502029].
  
  
    %\cite{Hawking:1982dh}
\bibitem{Hawking:1982dh} 
  S.~W.~Hawking and D.~N.~Page,
  ``Thermodynamics of Black Holes in anti-De Sitter Space,''
  Commun.\ Math.\ Phys.\  {\bf 87}, 577 (1983).
  %doi:10.1007/BF01208266
  %%CITATION = doi:10.1007/BF01208266;%%





\bibitem{curir1}
A.~Curir,
``Rotating black holes as dissipative spin-thermodynamical systems,''
Gen.\ Rel.\ Grav.\  {\bf 13}, 417 (1981).

\bibitem{curir2}
A.~Curir,
``Black hole emissions and phase transitions ,''
Gen.\ Rel.\ Grav.\  {\bf 13}, 1177 (1981).

%\cite{Pavon:1988in}
\bibitem{Pavon:1988in} 
  D.~Pavon and J.~M.~Rubi,
  ``Nonequilibrium Thermodynamic Fluctuations of Black Holes,''
  Phys.\ Rev.\ D {\bf 37}, 2052 (1988).
  
  %\cite{Pavon:1991kh}
\bibitem{Pavon:1991kh} 
  D.~Pavon,
  ``Phase transition in Reissner-Nordstrom black holes,''
  Phys.\ Rev.\ D {\bf 43}, 2495 (1991).

  \bibitem{Kaburakigrg}
O.~Kaburaki,
``Critical behavior of extremal Kerr-Newman black holes,''
Gen.\ Rel.\ Grav.\  {\bf 28}, 843 (1996).

%\cite{Cai:1996df}
\bibitem{Cai:1996df} 
  R.~G.~Cai, Z.~J.~Lu and Y.~Z.~Zhang,
  ``Critical behavior in (2+1)-dimensional black holes,''
  Phys.\ Rev.\ D {\bf 55}, 853 (1997)
  %doi:10.1103/PhysRevD.55.853
  [gr-qc/9702032].  
  
  %\cite{Cai:1998ep}
\bibitem{Cai:1998ep} 
  R.~G.~Cai and J.~H.~Cho,
  ``Thermodynamic curvature of the BTZ black hole,''
  Phys.\ Rev.\ D {\bf 60}, 067502 (1999)
  %doi:10.1103/PhysRevD.60.067502
  [hep-th/9803261].
  
  %\cite{Wei:2009zzf}
\bibitem{Wei:2009zzf} 
  Y.~H.~Wei,
  ``Thermodynamic critical and geometrical properties of charged BTZ black hole,''
  Phys.\ Rev.\ D {\bf 80}, 024029 (2009).
 % doi:10.1103/PhysRevD.80.024029

 %\cite{Bhattacharya:2019awq}
\bibitem{Bhattacharya:2019awq}
K.~Bhattacharya, S.~Dey, B.~R.~Majhi and S.~Samanta,
``General framework to study the extremal phase transition of black holes,''
Phys. Rev. D \textbf{99}, no.12, 124047 (2019)
%doi:10.1103/PhysRevD.99.124047
[arXiv:1903.03434 [gr-qc]].



   \bibitem{Kastor:2009wy} 
  D.~Kastor, S.~Ray and J.~Traschen,
  ``Enthalpy and the Mechanics of AdS Black Holes,''
  Class.\ Quant.\ Grav.\  {\bf 26}, 195011 (2009)
  %doi:10.1088/0264-9381/26/19/195011
  [arXiv:0904.2765 [hep-th]].
  %%CITATION = doi:10.1088/0264-9381/26/19/195011;%%
  
  %\cite{Dolan:2010ha}
\bibitem{Dolan:2010ha} 
  B.~P.~Dolan,
  ``The cosmological constant and the black hole equation of state,''
  Class.\ Quant.\ Grav.\  {\bf 28}, 125020 (2011)
  %doi:10.1088/0264-9381/28/12/125020
  [arXiv:1008.5023 [gr-qc]].
  %%CITATION = doi:10.1088/0264-9381/28/12/125020;%%
  
  %\cite{Dolan:2011xt}
\bibitem{Dolan:2011xt} 
  B.~P.~Dolan,
  ``Pressure and volume in the first law of black hole thermodynamics,''
  Class.\ Quant.\ Grav.\  {\bf 28}, 235017 (2011)
  %doi:10.1088/0264-9381/28/23/235017
  [arXiv:1106.6260 [gr-qc]].
  %%CITATION = doi:10.1088/0264-9381/28/23/235017;%%
  
  %\cite{Dolan:2011jm}
\bibitem{Dolan:2011jm} 
  B.~P.~Dolan,
  ``Compressibility of rotating black holes,''
  Phys.\ Rev.\ D {\bf 84}, 127503 (2011)
  %doi:10.1103/PhysRevD.84.127503
  [arXiv:1109.0198 [gr-qc]].
  %%CITATION = doi:10.1103/PhysRevD.84.127503;%%
  
  %\cite{Dolan:2012jh}
\bibitem{Dolan:2012jh} 
  B.~P.~Dolan,
  ``Where is the PdV term in the fist law of black hole thermodynamics?,''
  %doi:10.5772/52455
  arXiv:1209.1272 [gr-qc].
  %%CITATION = doi:10.5772/52455;%%
  
    %\cite{Kubiznak:2012wp}
\bibitem{Kubiznak:2012wp} 
  D.~Kubiznak and R.~B.~Mann,
  ``P-V criticality of charged AdS black holes,''
  JHEP {\bf 1207}, 033 (2012)
  %doi:10.1007/JHEP07(2012)033
  [arXiv:1205.0559 [hep-th]].
  %%CITATION = doi:10.1007/JHEP07(2012)033;%% 
  
    %\cite{Kubiznak:2016qmn}
\bibitem{Kubiznak:2016qmn} 
  D.~Kubiznak, R.~B.~Mann and M.~Teo,
  ``Black hole chemistry: thermodynamics with Lambda,''
  arXiv:1608.06147 [hep-th].
  %%CITATION = ARXIV:1608.06147;%% 

  %\cite{Bhattacharya:2017nru}
\bibitem{Bhattacharya:2017nru}
K.~Bhattacharya, B.~R.~Majhi and S.~Samanta,
``Van der Waals criticality in AdS black holes: a phenomenological study,''
Phys. Rev. D \textbf{96}, no.8, 084037 (2017)
%doi:10.1103/PhysRevD.96.084037
[arXiv:1709.02650 [gr-qc]].

%\cite{Bhattacharya:2021lgk}
\bibitem{Bhattacharya:2021lgk}
K.~Bhattacharya,
``Extended phase space thermodynamics of black holes: A study in Einstein's gravity and beyond,''
Nucl. Phys. B \textbf{989}, 116130 (2023)
%doi:10.1016/j.nuclphysb.2023.116130
[arXiv:2112.00938 [gr-qc]].






%\cite{Quevedo:2008ry}
\bibitem{Quevedo:2008ry}
H.~Quevedo and A.~Sanchez,
``Geometric description of BTZ black holes thermodynamics,''
Phys. Rev. D \textbf{79}, 024012 (2009)
%doi:10.1103/PhysRevD.79.024012
[arXiv:0811.2524 [gr-qc]].

%\cite{Akbar:2011qw}
\bibitem{Akbar:2011qw}
M.~Akbar, H.~Quevedo, K.~Saifullah, A.~Sanchez and S.~Taj,
``Thermodynamic Geometry Of Charged Rotating BTZ Black Holes,''
Phys. Rev. D \textbf{83}, 084031 (2011)
%doi:10.1103/PhysRevD.83.084031
[arXiv:1101.2722 [gr-qc]].

%\cite{Hendi:2015cka}
\bibitem{Hendi:2015cka}
S.~H.~Hendi and R.~Naderi,
``Geometrothermodynamics of black holes in Lovelock gravity with a nonlinear electrodynamics,''
Phys. Rev. D \textbf{91}, no.2, 024007 (2015)
%doi:10.1103/PhysRevD.91.024007
[arXiv:1510.06269 [hep-th]].

%\cite{Sarkar:2006tg}
\bibitem{Sarkar:2006tg}
T.~Sarkar, G.~Sengupta and B.~Nath Tiwari,
``On the thermodynamic geometry of BTZ black holes,''
JHEP \textbf{11}, 015 (2006)
%doi:10.1088/1126-6708/2006/11/015
[arXiv:hep-th/0606084 [hep-th]].

%\cite{Hendi:2015xya}
\bibitem{Hendi:2015xya}
S.~H.~Hendi, A.~Sheykhi, S.~Panahiyan and B.~Eslam Panah,
``Phase transition and thermodynamic geometry of Einstein-Maxwell-dilaton black holes,''
Phys. Rev. D \textbf{92}, no.6, 064028 (2015)
%doi:10.1103/PhysRevD.92.064028
[arXiv:1509.08593 [hep-th]].

%\cite{Banerjee:2016nse}
\bibitem{Banerjee:2016nse}
R.~Banerjee, B.~R.~Majhi and S.~Samanta,
``Thermogeometric phase transition in a unified framework,''
Phys. Lett. B \textbf{767}, 25-28 (2017)
%doi:10.1016/j.physletb.2017.01.040
[arXiv:1611.06701 [gr-qc]].

%\cite{Bhattacharya:2017hfj}
\bibitem{Bhattacharya:2017hfj}
K.~Bhattacharya and B.~R.~Majhi,
``Thermogeometric description of the van der Waals like phase transition in AdS black holes,''
Phys. Rev. D \textbf{95}, no.10, 104024 (2017)
%doi:10.1103/PhysRevD.95.104024
[arXiv:1702.07174 [gr-qc]].

%\cite{Bhattacharya:2019qxe}
\bibitem{Bhattacharya:2019qxe}
K.~Bhattacharya and B.~R.~Majhi,
``Thermogeometric study of van der Waals like phase transition in black holes: an alternative approach,''
Phys. Lett. B \textbf{802}, 135224 (2020)
%doi:10.1016/j.physletb.2020.135224
[arXiv:1903.10370 [gr-qc]].

%\cite{Cunha:2017qtt}
\bibitem{Cunha:2017qtt}
P.~V.~P.~Cunha, E.~Berti and C.~A.~R.~Herdeiro,
``Light-Ring Stability for Ultracompact Objects,''
Phys. Rev. Lett. \textbf{119}, no.25, 251102 (2017)
%doi:10.1103/PhysRevLett.119.251102
[arXiv:1708.04211 [gr-qc]].

%\cite{Cunha:2020azh}
\bibitem{Cunha:2020azh}
P.~V.~P.~Cunha and C.~A.~R.~Herdeiro,
``Stationary black holes and light rings,''
Phys. Rev. Lett. \textbf{124}, no.18, 181101 (2020)
%doi:10.1103/PhysRevLett.124.181101
[arXiv:2003.06445 [gr-qc]].

%\cite{Wei:2020rbh}
\bibitem{Wei:2020rbh}
S.~W.~Wei,
``Topological Charge and Black Hole Photon Spheres,''
Phys. Rev. D \textbf{102}, no.6, 064039 (2020)
%doi:10.1103/PhysRevD.102.064039
[arXiv:2006.02112 [gr-qc]].

%\cite{Guo:2020qwk}
\bibitem{Guo:2020qwk}
M.~Guo and S.~Gao,
``Universal Properties of Light Rings for Stationary Axisymmetric Spacetimes,''
Phys. Rev. D \textbf{103}, no.10, 104031 (2021)
%doi:10.1103/PhysRevD.103.104031
[arXiv:2011.02211 [gr-qc]].

%\cite{Cunha:2024ajc}
\bibitem{Cunha:2024ajc}
P.~V.~P.~Cunha, C.~A.~R.~Herdeiro and J.~P.~A.~Novo,
``Light rings on stationary axisymmetric spacetimes: blind to the topology and able to coexist,''
[arXiv:2401.05495 [gr-qc]].





%\cite{Wei:2022dzw}
\bibitem{Wei:2022dzw}
S.~W.~Wei, Y.~X.~Liu and R.~B.~Mann,
``Black Hole Solutions as Topological Thermodynamic Defects,''
Phys. Rev. Lett. \textbf{129}, no.19, 191101 (2022)
%doi:10.1103/PhysRevLett.129.191101
[arXiv:2208.01932 [gr-qc]].

%\cite{Wei:2021vdx}
\bibitem{Wei:2021vdx}
S.~W.~Wei and Y.~X.~Liu,
``Topology of black hole thermodynamics,''
Phys. Rev. D \textbf{105}, no.10, 104003 (2022)
%doi:10.1103/PhysRevD.105.104003
[arXiv:2112.01706 [gr-qc]].



%\cite{Yerra:2022alz}
\bibitem{Yerra:2022alz}
P.~K.~Yerra and C.~Bhamidipati,
``Topology of black hole thermodynamics in Gauss-Bonnet gravity,''
Phys. Rev. D \textbf{105}, no.10, 104053 (2022)
%doi:10.1103/PhysRevD.105.104053
[arXiv:2202.10288 [gr-qc]].

%\cite{Yerra:2022eov}
\bibitem{Yerra:2022eov}
P.~K.~Yerra and C.~Bhamidipati,
``Topology of Born-Infeld AdS black holes in 4D novel Einstein-Gauss-Bonnet gravity,''
Phys. Lett. B \textbf{835}, 137591 (2022)
%doi:10.1016/j.physletb.2022.137591
[arXiv:2207.10612 [gr-qc]].












%\cite{Gogoi:2023xzy}
\bibitem{Gogoi:2023xzy}
N.~J.~Gogoi and P.~Phukon,
``Thermodynamic topology of 4D dyonic AdS black holes in different ensembles,''
Phys. Rev. D \textbf{108}, no.6, 066016 (2023)
%doi:10.1103/PhysRevD.108.066016
[arXiv:2304.05695 [hep-th]].

%\cite{Gogoi:2023qku}
\bibitem{Gogoi:2023qku}
N.~J.~Gogoi and P.~Phukon,
``Topology of thermodynamics in R-charged black holes,''
Phys. Rev. D \textbf{107}, no.10, 106009 (2023)


%\cite{Yerra:2022coh}
\bibitem{Yerra:2022coh}
P.~K.~Yerra, C.~Bhamidipati and S.~Mukherji,
``Topology of critical points and Hawking-Page transition,''
Phys. Rev. D \textbf{106}, no.6, 064059 (2022)
%doi:10.1103/PhysRevD.106.064059
[arXiv:2208.06388 [hep-th]].

%\cite{Yerra:2023ocu}
\bibitem{Yerra:2023ocu}
P.~K.~Yerra, C.~Bhamidipati and S.~Mukherji,
``Topology of Hawking-Page transition in Born-Infeld AdS black holes,''
J. Phys. Conf. Ser. \textbf{2667}, no.1, 012031 (2023)
%doi:10.1088/1742-6596/2667/1/012031
[arXiv:2312.10784 [gr-qc]].

\bibitem{Yerra:2023} P.~K.~Yerra, C.~Bhamidipati and S.~Mukherji,  ``Topology of critical points in boundary matrix duals", [arXiv:2304.14988 [hep-th]]

%\cite{Barzi:2023msl}
\bibitem{Barzi:2023msl}
F.~Barzi, H.~El Moumni and K.~Masmar,
``R\'enyi Topology of Charged-flat Black Hole: Hawking-Page and Van-der-Waals Phase Transitions,''
[arXiv:2309.14069 [hep-th]].

%\cite{Ahmed:2022kyv}
\bibitem{Ahmed:2022kyv}
M.~B.~Ahmed, D.~Kubiznak and R.~B.~Mann,
``Vortex-antivortex pair creation in black hole thermodynamics,''
Phys. Rev. D \textbf{107}, no.4, 046013 (2023)
%doi:10.1103/PhysRevD.107.046013
[arXiv:2207.02147 [hep-th]].

%\cite{Wei:2022mzv}
\bibitem{Wei:2022mzv}
S.~W.~Wei and Y.~X.~Liu,
``Topology of equatorial timelike circular orbits around stationary black holes,''
Phys. Rev. D \textbf{107}, no.6, 064006 (2023)
%doi:10.1103/PhysRevD.107.064006
[arXiv:2207.08397 [gr-qc]].

%\cite{Fan:2022bsq}
\bibitem{Fan:2022bsq}
Z.~Y.~Fan,
``Topological interpretation for phase transitions of black holes,''
Phys. Rev. D \textbf{107}, no.4, 044026 (2023)
%doi:10.1103/PhysRevD.107.044026
[arXiv:2211.12957 [gr-qc]].

%\cite{Wu:2022whe}
\bibitem{Wu:2022whe}
D.~Wu,
``Topological classes of rotating black holes,''
Phys. Rev. D \textbf{107}, no.2, 024024 (2023)
%doi:10.1103/PhysRevD.107.024024
[arXiv:2211.15151 [gr-qc]].

%\cite{Fang:2022rsb}
\bibitem{Fang:2022rsb}
C.~Fang, J.~Jiang and M.~Zhang,
``Revisiting thermodynamic topologies of black holes,''
JHEP \textbf{01}, 102 (2023)
%doi:10.1007/JHEP01(2023)102
[arXiv:2211.15534 [gr-qc]].

%\cite{Wu:2023xpq}
\bibitem{Wu:2023xpq}
D.~Wu,
``Classifying topology of consistent thermodynamics of the four-dimensional neutral Lorentzian NUT-charged spacetimes,''
Eur. Phys. J. C \textbf{83}, no.5, 365 (2023)
%doi:10.1140/epjc/s10052-023-11561-4
[arXiv:2302.01100 [gr-qc]].

\bibitem{Wu:ejpc2023} D.~Wu, ``Consistent thermodynamics and topological classes for the four-dimensional Lorentzian charged Taub-NUT spacetimes", Eur. Phys. J. C {\bf 83}, 589 (2023)
[arXiv: 2306.02324 [gr-qc]].

%\cite{Wu:2023sue}
\bibitem{Wu:2023sue}
D.~Wu and S.~Q.~Wu,
``Topological classes of thermodynamics of rotating AdS black holes,''
Phys. Rev. D \textbf{107}, no.8, 084002 (2023)
%doi:10.1103/PhysRevD.107.084002
[arXiv:2301.03002 [hep-th]].

%\cite{Li:2023ppc}
\bibitem{Li:2023ppc}
R.~Li, C.~Liu, K.~Zhang and J.~Wang,
``Topology of the landscape and dominant kinetic path for the thermodynamic phase transition of the charged Gauss-Bonnet-AdS black holes,''
Phys. Rev. D \textbf{108}, no.4, 044003 (2023)
%doi:10.1103/PhysRevD.108.044003
[arXiv:2302.06201 [gr-qc]].

%\cite{Wei:2023bgp}
\bibitem{Wei:2023bgp}
S.~W.~Wei, Y.~P.~Zhang, Y.~X.~Liu and R.~B.~Mann,
``Static spheres around spherically symmetric black hole spacetime,''
Phys. Rev. Res. \textbf{5}, no.4, 043050 (2023)
%doi:10.1103/PhysRevResearch.5.043050
[arXiv:2303.06814 [gr-qc]].

%\cite{Alipour:2023uzo}
\bibitem{Alipour:2023uzo}
M.~R.~Alipour, M.~A.~S.~Afshar, S.~Noori Gashti and J.~Sadeghi,
``Topological classification and black hole thermodynamics,''
Phys. Dark Univ. \textbf{42}, 101361 (2023)
%doi:10.1016/j.dark.2023.101361
[arXiv:2305.05595 [gr-qc]].

%\cite{Zhang:2023uay}
\bibitem{Zhang:2023uay}
M.~Zhang and J.~Jiang,
``Bulk-boundary thermodynamic equivalence: a topology viewpoint,''
JHEP \textbf{06}, 115 (2023)
%doi:10.1007/JHEP06(2023)115
[arXiv:2303.17515 [hep-th]].

%\cite{Sadeghi:2023aii}
\bibitem{Sadeghi:2023aii}
J.~Sadeghi, S.~Noori Gashti, M.~R.~Alipour and M.~A.~S.~Afshar,
``Bardeen black hole thermodynamics from topological perspective,''
Annals Phys. \textbf{455}, 169391 (2023)
%doi:10.1016/j.aop.2023.169391
[arXiv:2306.05692 [hep-th]].

\bibitem{Sageghi-2023}
J.~Sadeghi, M.~A.~S.~Afshar, S.~Noori Gashti, M.~R.~Alipour and ,
``Thermodynamic topology of black holes from bulk-boundary, extended, and restricted phase space perspectives", Annals Phys. {\bf 460}, 169569 (2024)
[arXiv:2312.04325 [hep-th]].

%\cite{Liu:2023sbf}
\bibitem{Liu:2023sbf}
C.~Liu, R.~Li, K.~Zhang and J.~Wang,
``Generalized free energy and dynamical state transition of the dyonic AdS black hole in the grand canonical ensemble,''
JHEP \textbf{11}, 068 (2023)
%doi:10.1007/JHEP11(2023)068
[arXiv:2309.13931 [gr-qc]].

\bibitem{Bai-2023}
N.~C.~Bai, L.~Li and J~Tao,
``Topology of black hole thermodynamics in Lovelock gravity",
Phys. Rev. D {\bf 107}, 064015 (2023)
[arXiv:2208.10177 [gr-qc]]

\bibitem{edery-2021} A. Edery, ``Non-singular vortices with positive mass in 2+1-dimensional Einstein gravity with AdS$_3$ and Minkowski background",
JHEP {\bf 01}, 166 (2021) 
[arXiv:2004.0925 [hep-th]].
 

\bibitem{edery-2022} A. Edery, ``Nonminimally coupled gravitating vortex: phase transition at critical coupling $\xi_c $ in AdS$_3$,"
Phys. Rev. D {\bf 106}, 065017 (2022) 
[arXiv:2205.12175 [hep-th]].

%------------papers suggested by referee------------


%\cite{Lei:2024wvj}
\bibitem{Lei:2024wvj}
C.~Lei, Y.~Liu and D.~Chen,
``Topological properties of black rings,''
Nucl. Phys. B \textbf{1002}, 116527 (2024).

%\cite{Shahzad:2024ojx}
\bibitem{Shahzad:2024ojx}
M.~U.~Shahzad, A.~Mehmood, A.~Malik and A.~\"Ovg\"un,
``Topological behavior of 3D regular black hole with zero point length,''
Phys. Dark Univ. \textbf{44}, 101437 (2024).

%\cite{Shahzad:2023cis}
\bibitem{Shahzad:2023cis}
M.~U.~Shahzad, A.~Mehmood, S.~Sharif and A.~\"Ovg\"un,
``Criticality and topological classes of neutral Gauss\textendash{}Bonnet AdS black holes in 5D,''
Annals Phys. \textbf{458}, no.3, 169486 (2023)
%doi:10.1016/j.aop.2023.169486

%\cite{Wu:2024rmv}
\bibitem{Wu:2024rmv}
D.~Wu, S.~Y.~Gu, X.~D.~Zhu, Q.~Q.~Jiang and S.~Z.~Yang,
``Topological classes of thermodynamics of the static multi-charge AdS black holes in gauged supergravities,''
[arXiv:2402.00106 [hep-th]].

%------------papers suggested by referee------------


%\cite{Duan:1979ucg}
\bibitem{Duan:1979ucg}
Y.~S.~Duan and M.~L.~Ge,
``SU(2) Gauge Theory and Electrodynamics with N Magnetic Monopoles,''
Sci. Sin. \textbf{9}, no.11, 1072 (1979).

\bibitem{duan1}
Y.~S.~Duan, ``The structure of the topological current'',
SLAC-PUB-3301, (1984).

\bibitem{goursat}
E. Goursat, \textit{``A Course in Mathematical Analysis''}, translated by E.~R.~Hedrick (Dover, New York, 1904), Vol. I.

%\cite{Duan:1998kw}
\bibitem{Duan:1998kw}
Y.~S.~Duan, S.~Li and G.~H.~Yang,
``The bifurcation theory of the Gauss-Bonnet-Chern topological current and Morse function,''
Nucl. Phys. B \textbf{514}, 705-720 (1998)~.

\bibitem{A.S.Schwarz}
A.S.Schwarz, \textit{“Topology for physicists”}, DOI: 	10.1007/978-3-662-02998-5

%\cite{Jardim:2012se}
\bibitem{Jardim:2012se}
D.~F.~Jardim, M.~E.~Rodrigues and M.~J.~S.~Houndjo,
``Thermodynamics of phantom Reissner-Nordstrom-AdS black hole,''
Eur. Phys. J. Plus \textbf{127}, 123 (2012)
%doi:10.1140/epjp/i2012-12123-x
[arXiv:1202.2830 [gr-qc]].

%\cite{Banados:1992wn}
\bibitem{Banados:1992wn}
M.~Banados, C.~Teitelboim and J.~Zanelli,
``The Black hole in three-dimensional space-time,''
Phys. Rev. Lett. \textbf{69}, 1849-1851 (1992)
%doi:10.1103/PhysRevLett.69.1849
[arXiv:hep-th/9204099 [hep-th]].

%\cite{Banados:1992gq}
\bibitem{Banados:1992gq}
M.~Banados, M.~Henneaux, C.~Teitelboim and J.~Zanelli,
``Geometry of the (2+1) black hole,''
Phys. Rev. D \textbf{48}, 1506-1525 (1993)
[erratum: Phys. Rev. D \textbf{88}, 069902 (2013)]
%doi:10.1103/PhysRevD.48.1506
[arXiv:gr-qc/9302012 [gr-qc]].

%\cite{Eling:2006aw}
\bibitem{Eling:2006aw}
C.~Eling, R.~Guedens and T.~Jacobson,
``Non-equilibrium thermodynamics of spacetime,''
Phys. Rev. Lett. \textbf{96}, 121301 (2006)
%doi:10.1103/PhysRevLett.96.121301
[arXiv:gr-qc/0602001 [gr-qc]].

%\cite{Elizalde:2008pv}
\bibitem{Elizalde:2008pv}
E.~Elizalde and P.~J.~Silva,
``F(R) gravity equation of state,''
Phys. Rev. D \textbf{78}, 061501 (2008)
%doi:10.1103/PhysRevD.78.061501
[arXiv:0804.3721 [hep-th]].

%\cite{Chirco:2009dc}
\bibitem{Chirco:2009dc}
G.~Chirco and S.~Liberati,
``Non-equilibrium Thermodynamics of Spacetime: The Role of Gravitational Dissipation,''
Phys. Rev. D \textbf{81}, 024016 (2010)
%doi:10.1103/PhysRevD.81.024016
[arXiv:0909.4194 [gr-qc]].

%\cite{Padmanabhan:2002sha}
\bibitem{Padmanabhan:2002sha}
T.~Padmanabhan,
``Classical and quantum thermodynamics of horizons in spherically symmetric space-times,''
Class. Quant. Grav. \textbf{19}, 5387-5408 (2002)
%doi:10.1088/0264-9381/19/21/306
[arXiv:gr-qc/0204019 [gr-qc]].

  %\cite{Padmanabhan:2003gd}
\bibitem{Padmanabhan:2003gd}
T.~Padmanabhan,
``Gravity and the thermodynamics of horizons,''
Phys. Rept. \textbf{406}, 49-125 (2005)
%doi:10.1016/j.physrep.2004.10.003
[arXiv:gr-qc/0311036 [gr-qc]].

%\cite{Bamba:2009gq}
\bibitem{Bamba:2009gq}
K.~Bamba, C.~Q.~Geng, S.~Nojiri and S.~D.~Odintsov,
``Equivalence of modified gravity equation to the Clausius relation,''
EPL \textbf{89}, no.5, 50003 (2010)
%doi:10.1209/0295-5075/89/50003
[arXiv:0909.4397 [hep-th]].

%\cite{Faraoni:2010yi}
\bibitem{Faraoni:2010yi} 
V.~Faraoni,
``Black hole entropy in scalar-tensor and f(R) gravity: An Overview,''
Entropy {\bf 12}, 1246 (2010)
%doi:10.3390/e12051246
[arXiv:1005.2327 [gr-qc]].
%%CITATION = doi:10.3390/e12051246;%%

%\cite{Jacobson:1993vj}
\bibitem{Jacobson:1993vj}
T.~Jacobson, G.~Kang and R.~C.~Myers,
``On black hole entropy,''
Phys. Rev. D \textbf{49}, 6587-6598 (1994)
%doi:10.1103/PhysRevD.49.6587
[arXiv:gr-qc/9312023 [gr-qc]].

%\cite{Iyer:1994ys}
\bibitem{Iyer:1994ys}
V.~Iyer and R.~M.~Wald,
``Some properties of Noether charge and a proposal for dynamical black hole entropy,''
Phys. Rev. D \textbf{50}, 846-864 (1994)
%doi:10.1103/PhysRevD.50.846
[arXiv:gr-qc/9403028 [gr-qc]].

%\cite{Visser:1993nu}
\bibitem{Visser:1993nu}
M.~Visser,
``Dirty black holes: Entropy as a surface term,''
Phys. Rev. D \textbf{48}, 5697-5705 (1993)
%doi:10.1103/PhysRevD.48.5697
[arXiv:hep-th/9307194 [hep-th]].

%\cite{Bhattacharya:2022mnb}
\bibitem{Bhattacharya:2022mnb}
K.~Bhattacharya and B.~R.~Majhi,
``Scalar\textendash{}tensor gravity from thermodynamic and fluid-gravity perspective,''
Gen. Rel. Grav. \textbf{54}, no.9, 112 (2022)
%doi:10.1007/s10714-022-02999-0
[arXiv:2209.07050 [gr-qc]].








%-----PV criticality in f(R) gravity

%\cite{Chen:2013ce}
\bibitem{Chen:2013ce}
S.~Chen, X.~Liu, C.~Liu and J.~Jing,
``$P-V$ criticality of AdS black hole in $f(R)$ gravity,''
Chin. Phys. Lett. \textbf{30}, 060401 (2013)
%doi:10.1088/0256-307X/30/6/060401
[arXiv:1301.3234 [gr-qc]].

%\cite{Ovgun:2017bgx}
\bibitem{Ovgun:2017bgx}
A.~\"Ovg\"un,
``$P-v$ criticality of a specific black hole in $f(R)$ gravity coupled with Yang-Mills field,''
Adv. High Energy Phys. \textbf{2018}, 8153721 (2018)
%doi:10.1155/2018/8153721
[arXiv:1710.06795 [gr-qc]].


%------in brans dicke theory

%\cite{Hendi:2015kza}
\bibitem{Hendi:2015kza}
S.~H.~Hendi and Z.~Armanfard,
``Extended phase space thermodynamics and $P-V$ criticality of charged black holes in Brans\textendash{}Dicke theory,''
Gen. Rel. Grav. \textbf{47}, no.10, 125 (2015)
%doi:10.1007/s10714-015-1965-6
[arXiv:1503.07070 [gr-qc]]. 

%\cite{Hendi:2015hgg}
\bibitem{Hendi:2015hgg}
S.~H.~Hendi, R.~M.~Tad, Z.~Armanfard and M.~S.~Talezadeh,
``Extended phase space thermodynamics and P\textendash{}V criticality: Brans\textendash{}Dicke\textendash{}Born\textendash{}Infeld vs. Einstein\textendash{}Born\textendash{}Infeld-dilaton black holes,''
Eur. Phys. J. C \textbf{76}, no.5, 263 (2016)
%doi:10.1140/epjc/s10052-016-4106-9
[arXiv:1511.02761 [gr-qc]].
%---------------------------------------

%\cite{Majhi:2012fz}
\bibitem{Majhi:2012fz}
B.~R.~Majhi and D.~Roychowdhury,
``Phase transition and scaling behavior of topological charged black holes in Horava-Lifshitz gravity,''
Class. Quant. Grav. \textbf{29}, 245012 (2012)
%doi:10.1088/0264-9381/29/24/245012
[arXiv:1205.0146 [gr-qc]].

\bibitem{Du:2019poh}
Y.~Z.~Du, H.~H.~Zhao and L.~C.~Zhang,
``Microstructure and Continuous Phase Transition of the Einstein-Gauss-Bonnet AdS Black Hole,''
Adv. High Energy Phys. \textbf{2020}, 6395747 (2020)
%doi:10.1155/2020/6395747
[arXiv:1901.07932 [hep-th]].

\bibitem{Ali:2023zgm}
A.~Ali,
``Topologically nontrivial black holes of Lovelock gravity sourced by logarithmic electrodynamics,''
Eur. Phys. J. C \textbf{83}, no.7, 624 (2023)


%\cite{EslamPanah:2024tex}
\bibitem{EslamPanah:2024tex}
B.~Eslam Panah,
``Analytic Electrically Charged Black Holes in F(R)-ModMax Theory,''
PTEP \textbf{2024}, no.2, 023E01 (2024)
%doi:10.1093/ptep/ptae012
[arXiv:2402.12492 [gr-qc]].

%\cite{Ayuso:2020rmb}
\bibitem{Ayuso:2020rmb}
I.~Ayuso and D.~Saez-Chillon Gomez,
``Extremal cosmological black holes in Horndeski gravity and the anti-evaporation regime,''
Universe \textbf{6}, no.11, 210 (2020)
%doi:10.3390/universe6110210
[arXiv:2004.10139 [gr-qc]].

\bibitem{TranNHung:2024pig}
T.~N.~Hung and C.~H.~Nam,
``Generalized free energy and thermodynamic phases of black holes in the gauged Kaluza-Klein theory,''
[arXiv:2403.08322 [gr-qc]].










\end{thebibliography}
\end{document}